\documentclass[english,prc,aps,superscriptaddress]{revtex4}
\usepackage{graphicx}
\usepackage{amsmath}
\usepackage{amssymb}

\makeatletter

\usepackage{babel}
\makeatother

\begin{document}

\title{Nuclear Schiff moment and soft vibrational modes}

\author{Vladimir Zelevinsky}
\affiliation{Department of Physics and Astronomy Michigan State University, East Lansing, MI
48824-1321, USA}
\affiliation{National Superconducting Cyclotron Laboratory, Michigan State University, East Lansing, MI 48824-1321, USA}

\author{Alexander Volya}
\affiliation{Department of Physics, Florida State University, Tallahassee, FL
32306-4350, USA}

\author{Naftali Auerbach}
\affiliation{Department of Physics and Astronomy Michigan State University, East Lansing, MI
48824-1321, USA}
\affiliation{School of Physics and Astronomy, Tel Aviv University, Tel Aviv, 69978,
Israel}

\begin{abstract}
The atomic electric dipole moment (EDM) currently searched by a
number of experimental groups requires that both parity and
time-reversal invariance be violated. According to current
theoretical understanding, the EDM is induced by the nuclear Schiff
moment. The enhancement of the Schiff moment by the combination of
static quadrupole and octupole deformation was predicted earlier.
Here we study a further idea of the possible enhancement in the
absence of static deformation but in a nuclear system with soft
collective vibrations of two types. Both analytical approximation
and numerical solution of the simplified problem confirm the
presence of the enhancement. We discuss related aspects of nuclear
structure which should be studied beyond mean-field and random phase
approximations.

\end{abstract}
\maketitle

\section{Introduction}

Efforts of many experimental groups are directed to the discovery
of the \textsl{electric dipole moment} (EDM) of atoms. The existence
of the EDM in a stationary state of a finite quantum system would
manifest the \textsl{simultaneous violation} of parity (${\cal P}$-)
and time-reversal (${\cal T}$-) invariance \citep{purcell50}. The
best current experimental limits \citep{jacobs95,romalis01} are found
for $^{129}$Xe and $^{199}$Hg. Further progress in this field can
be connected with the better understanding of nuclear many-body mechanisms
which could substantially enhance the effect. This problem is also
of significant interest for theory of collective motion in nuclei
and other mesoscopic systems.

The atomic dipole moment is induced by the weak interactions of atomic
electrons with the nucleus. The dipole moment of the nucleus is almost
completely screened by the redistribution of atomic electrons in the
applied external electric field (the famous \textsl{Schiff theorem}
\citep{schiff63}). Instead of the nuclear dipole moment, one needs
to consider the vector operator of the next order, the so-called \textsl{Schiff
moment} \citep{FKS84},
\begin{equation}
{\bf S}=\frac{1}{10}\,\sum_{a}e_{a}{\bf r}_{a}\left[r_{a}^{2}-
\frac{5}{3}\langle r_{{\rm ch}}^{2}\rangle\right].     \label{ 1}
\end{equation}
The correctness of this conclusion was confirmed in the recent
theoretical discussion that lead to the reexamination of the Schiff
theorem \citep{liu07,senkov08}. The similar operator is known as
generating the isoscalar giant dipole resonance in nuclei since the
isoscalar dipole moment creates only a spurious mode of the
center-of-mass displacement.

In the presence of ${\cal P,T}$-violating weak interaction, the
non-zero expectation value of the Schiff moment, as of any polar
time-even vector, becomes possible in the ground state of a nucleus
with non-zero angular momentum $I$,
\begin{equation}
\langle{\bf S}\rangle=\langle({\bf S}\cdot{\bf I})\rangle\,
\frac{{\bf I}}{I(I+1)}\cdot                \label{2}
\end{equation}
In the next step, this vector induces the EDM of the atom that is
enhanced in heavy atoms. Typically, the Schiff moment has a
single-particle order of magnitude value as confirmed by various
calculations of nuclear structure, see for example
\citep{FV94,DS03,DSA05,jesus05}; the core polarization effects can
increase the result by a factor of order 2.

It is however known that there exist promising collective mechanisms
which must be explored in the search of the many-body enhancement of
the Schiff moment. Parity non-conservation in scattering of
polarized neutrons off heavy nuclei is enhanced by orders of
magnitude \citep{flam95} because of extreme proximity of $p$- and
$s$-wave resonances and the \textsl{chaotic nature} of their
extremely complicated wave functions. In contrast, here we are
interested in \textsl{coherent} mechanisms which could show up in
the properties of the ground states and low-lying excitations.
Essentially we need the collective states of opposite parity in the
vicinity of the ground state which can be effectively admixed to the
ground state by the weak interaction. The main hope here is
associated with \textsl{coexistence of quadrupole and octupole
collective degrees of freedom}. Indeed, it was shown that the
\textsl{intrinsic} Schiff moment (in the body-fixed frame), that
exists without violation of fundamental symmetries, can be enhanced
by 2-3 orders of magnitude in nuclei with simultaneous quadrupole
and octupole static deformation \citep{AFS96,SAF97,engel03}. The
mixing by the weak interaction of states with certain angular
momentum in the laboratory (space-fixed) frame will be particularly
enhanced for \textsl{parity doublets}
\citep{auerbach94,auerbach95,FZ95}. Those are the states with same
spin and opposite parity which differ by the {}``right\char`\"{} and
{}``left\char`\"{} orientation of the asymmetric configuration that
is characteristic for the deformation without reflection symmetry in
the equatorial plane.

The requirements of static octupole deformation can be relaxed since
almost the same effect can be reached \citep{engel00,FZ95} when
static quadrupole deformation is combined with a soft octupole
vibrational mode. Similarly to the case of static octupole
deformation, where the enhancement effect is proportional to the
square $\beta_{3}^{2}$ of the octupole deformation parameter, here
the analogous result contains the dynamic mean square value
$\langle\beta_{3}^{2}\rangle$ that has the same order of magnitude
being proportional to the inverse soft mode frequency,
$1/\omega_{3}$. This significantly widens the set of nuclear
candidates where one can expect the collective enhancement of the
Schiff moment.

There is a theoretical possibility that the enhancement may occur
even in the absence of static quadrupole deformation due to the
combination of \textsl{quadrupole and octupole soft modes}
\citep{FZ03,ADFLSZ06}. This would broaden even further the
possibilities for the experimental search of the EDM adding, for
example, such nuclei as light isotopes of radium and radon, where
the presence of both soft modes with strong dipole transitions
connecting corresponding vibrational bands is well known
\citep{cocks97}. Although phenomenological estimates \citep{FZ03}
supported this idea, the detailed microscopic calculation
\citep{ADFLSZ06} based on the random phase approximation (RPA) and
quasiparticle-phonon coupling gave essentially negative results. The
effect of strong enhancement was found only in the unphysical limit
of very low vibrational frequencies $\omega_{2}$ and $\omega_{3}$.
However, the applicability of the RPA to the situation of strong
interaction of low-frequency collective modes is questionable and
that result might be an artifact of the inappropriate approximation.

In the current work we present a simplified model, where an unpaired
particle interacts with two coupled soft collective modes. This
interaction is effectively strong and creates a \textsl{coherent
state (condensate) of phonons} although the even core is still
spherical. Such a situation cannot be adequately treated in the RPA
framework. We consider both the approximate analytical approach and
exact numerical solution. The results clearly demonstrate the
existence of a parameter region, where we see the strong enhancement
of the nuclear Schiff moment. On the other hand, the whole problem
is quite interesting irrespectively of the symmetry violations, for
the study of nuclear structure in the situation of shape instability
due to the presence of soft modes. A more detailed review of the
entire problem of the Schiff moment and the previous search for its
collective enhancement can be found in \citep{AZ08}.

\section{Microscopic justification of enhancement}

\subsection{Spontaneous symmetry breaking in an even nucleus with soft modes}

We consider a generic model of a {}``soft\char`\"{} nucleus that has
well developed vibrational modes still keeping, on average, a
spherical shape. Although the non-zero Schiff moment requires
non-zero nuclear spin and therefore an odd-$A$ nucleus, we start
with the neighboring even-even nuclei. Here we assume that the
collective modes with spin-parity characteristics $2^{+}$ and
$3^{-}$ have low excitation energies $\omega_{2}$ and $\omega_{3}$,
well below the gap energy $2\Delta$ that separates the ground state
$0^{+}$ from the threshold of two-quasiparticle states. This
situation, for example, can be found in the chain of even-even xenon
isotopes \citep{mueller06} and, most clearly, in light isotopes of
radon and radium \citep{cocks97}, where we see soft quadrupole and
octupole collective modes evidently correlated in their energetics.
As the energy spectrum does not display a rotational pattern, we can
assume that these nuclei are still spherical. However, the spectra
do not show usual phonon multiplets either. In radium and radon
isotopes, strong dipole transitions are present between the
corresponding members of long quasivibrational bands of positive and
negative parity. This situation seems to be favorable for dipole
correlations, and consequently for the Schiff moment.

The correlation between quadrupole and octupole vibrations is an
anharmonic effect that can be theoretically described only if one
goes beyond the random phase approximation (RPA). In the following
part of the article, Sec. III, we show a numerically solved
two-level model that starts from single-particle mean field levels
and an inter-particle interaction; then both phonons and anharmonic
effects result from the exact diagonalization. In the present
Section we use a simpler analytical approximation of bosonic phonons
interacting with unpaired particles. The model is a generalization
of the approach suggested in Ref. \citep{metlay95} for a global
analysis of quadrupole-octupole correlations. This idea was later
used for the explanation of such correlations seen in the
experimental data for xenon isotopes \citep{mueller06} and in the
schematic arguments applied to the ehnancement of the Schiff moment
in Ref. \citep{ADFLSZ06}. The correlations between quadrupole and
octupole modes can lead to new observable effects and phase
transitions also in nuclei with $A>220$ as it was discussed recently
\citep{frauendorf08,bizzeti08}.

The model includes two groups of spherical single-particle levels
with spins $\{j\}$ and $\{j'\}$ of positive and negative parity,
respectively. The elementary bosonic quanta, phonons, are coherent
superpositions of many two-quasiparticle excitations. The quadrupole
phonons $d_{\mu}$ contain $(j_{1}j_{2})_{2\mu}$ and
$(j_{1}'j_{2}')_{2\mu}$ combinations allowed by the angular momentum
coupling, while the octupole phonons $f_{\mu}$ are built of the
pairs $(j_{1}j_{2}')_{3\mu}$ based on the orbitals of opposite
parity. The effective Hamiltonian of low-lying states in the even
nucleus is given by
\begin{equation}
H=H_{2}+H_{3}+H_{23},                             \label{3}
\end{equation}
where $H_{2}$ and $H_{3}$ describe quadrupole and octupole
collective modes (in principle including their anharmonicity). The
interaction term,
\begin{equation}
H_{23}=x\sum_{\mu}\Bigl[(f^{\dagger}f)_{2\mu}d_{\mu}^{\dagger}+ {\rm
h.c.}\Bigr],                                   \label{4}
\end{equation}
is responsible for the mode-mode interaction with $x$ as a coupling
constant. The construction of (\ref{4}) is the simplest one that
accounts for parity and angular momentum conservation.

The non-linear interaction (\ref{4}) leads to the effective
deformation of the excited states when both modes are simultaneously
present. Indeed, the low-lying quadrupole and octupole modes support
each other and generate the \textsl{coherent state} with the
non-vanishing expectation value of the quadrupole moment according
to
\begin{equation}
\langle d_{\mu}\rangle=-\frac{x}{\omega_{2}}\,
\langle(f^{\dagger}f)_{2\mu}\rangle,              \label{5}
\end{equation}
as follows from the operator equation of motion
\begin{equation}
[d_{\mu},H]=\omega_{2}d_{\mu}+x(f^{\dagger}f)_{2\mu}.   \label{6}
\end{equation}
for the quadrupole phonon operator $d_{\mu}$. For simplicity, here
we neglect the possible quadrupole anharmonicity that would only
substitute the unperturbed quadrupole frequency $\omega_{2}$ by the
effective curvature of the ground state quadrupole potential
\citep{vorov83}.

The result (\ref{5}) means \textsl{spontaneous violation of rotational
symmetry} by the soft quadrupole mode that is determined by the octupole
phonon which selects the intrinsic axis (we assume axial symmetry).
The direction of the effective self-consistent deformation is arbitrary,
and, to restore the symmetry and appropriate quantum numbers of total
nuclear spin in the space-fixed coordinate frame, we accept that the
orientation of the deformation is given by a spherical function of
corresponding rank,
\begin{equation}
\langle d_{\mu}\rangle=\delta_{2}\,\sqrt{\frac{4\pi}{5}}\,
Y_{2\mu}^{\ast}({\bf n}),\quad\langle f_{\mu}\rangle=\hat{f}\,
\sqrt{\frac{4\pi}{7}}\, Y_{3\mu}^{\ast}({\bf n}).   \label{7}
\end{equation}
Here ${\bf n}$ is the unit vector of the symmetry axis considered as
a variable in the collective space. A similar operator approach was
used long ago in the derivation of the nuclear moment of inertia
without applying a cranking model \citep{BZ70}.

The number $N_{3}=\sum_{\mu}f_{\mu}^{\dagger}f_{\mu}=\hat{f}^{\dagger}\hat{f}$
of octupole phonons is conserved, $N_{3}=1$ in the lowest $3^{-}$
state, and $\hat{f}^{\dagger}$ is the operator generating the octupole
vibrational mode in the body-fixed frame defined by the orientation
${\bf n}$. Then eq. (\ref{5}) equates the ${\bf n}$-dependence
and, with the ansatz (\ref{7}), provides the effective quadrupole
deformation parameter $\delta_{2}$,
\begin{equation}
\delta_{2}=-\frac{x}{\omega_{2}}\,\sqrt{5}\left(\begin{array}{ccc}
3 & 3 & 2\\
0 & 0 & 0\end{array}\right)=-\sqrt{\frac{4}{21}}\,\frac{x}{\omega_{2}}.
                                                     \label{8}
\end{equation}
The $1/\omega$-dependence of the effective deformation parameters is
the characteristic feature of the whole consideration based on the
assumption of soft collective modes.

The equation of motion for the octupole mode, given by the
commutator $[f_{\mu},H]$ and the effective quadrupole parameter
(\ref{8}), is linear. It relates the excitation energy $E_{3}$ of
the octupole phonon with the corresponding unperturbed energy
$\omega_{3}$ and the quadrupole condensate (\ref{8}). Collecting
again the terms expressing the angular dependence, we obtain
\begin{equation}
E_{3}=\omega_{3}-\frac{8}{21}\,\frac{x^{2}}{\omega_{2}}.\label{9}
\end{equation}
This simple regularity first discussed in Ref. \citep{metlay95} in
the global review of octupole vibrations provides a clear
correlation between the two modes. Recent measurements for the long
chain of even-even xenon isotopes \citep{mueller06} show precisely
such a correlation, with a rather large magnitude for the parameter
$x$ that exceeds the expectations for the anharmonic mode-mode
coupling based on the standard RPA estimates. We assume that the
Schiff vector ${\bf S}$ has a non-zero reduced matrix element
$S^{\circ}$ between the one-phonon states of the two modes,
$|2^{+}\rangle$ and $|3^{-}\rangle$.

\subsection{Phonon condensate in an odd nucleus}

Now we consider the situation in the adjacent odd-$A$ nucleus, where
the presence of the unpaired particle in the ground state breaks symmetry
for both soft modes. The state of the odd particle is defined by the
density matrix
\begin{equation}
\rho_{j_{1}m_{1};j_{2}m_{2}}=a_{j_{2}m_{2}}^{\dagger}a_{j_{1}m_{1}}\label{10}
\end{equation}
in terms of the operators of creation, $a_{jm}^{\dagger}$, and
annihilation, $a_{jm}$, of the particle. The expectation value
$\langle\rho_{j_{1}m_{1};j_{2}m_{2}}\rangle$ in the ground state
with spontaneously broken symmetry can be written as
\begin{equation}
\langle\rho_{j_{1}m_{1};j_{2}m_{2}}\rangle=
\sum_{L\Lambda}(-)^{L-\Lambda+j_{2}-m_{2}}\left(\begin{array}{ccc}
j_{1} & L & j_{2}\\
m_{1} & -\Lambda & -m_{2}\end{array}\right)\rho_{L}(j_{1}j_{2})
Y_{L\Lambda}^{\ast}({\bf n}).                  \label{11}
\end{equation}
Here the even-$L$ parts come from the pairs of levels
$(j_{1},j_{2})$ or $(j_{1}',j_{2}')$ of the same parity, whereas the
odd-$L$ ones correspond to the combinations $(j_{1},j_{2}')$ and
$(j_{1}',j_{2})$ of single-particle levels of opposite parity.

In the odd nucleus, the unpaired particle interacts with the soft
modes, and the phonon Hamiltonian should be supplemented with
\begin{equation}
H_{{\rm odd}}=h+H^{(+)}+H'^{(+)}+H^{(-)},              \label{12}
\end{equation}
where $h$ contains unperturbed spherical single-particle energies
$\epsilon_{j}$ and $\epsilon_{j'}$, or quasiparticle energies if
pairing is included. The positive-parity parts are given by
\[H^{(+)}=-\sum_{j_{1}j_{2}m_{1}m_{2}\mu}x_{j_{1}j_{2}}^{(+)}
\rho_{j_{1}m_{1};j_{2}m_{2}}\]
\begin{equation}
\times\Bigl(d_{\mu}^{\dagger}+(-)^{\mu}d_{-\mu}\Bigr)
(-)^{\mu+j_{2}-m_{2}}\left(\begin{array}{ccc}
j_{1} & 2 & j_{2}\\
m_{1} & -\mu & -m_{2}\end{array}\right),           \label{13}
\end{equation}
and $H'^{(+)}$, where all positive parity levels $j$ are substituted
by the negative parity levels $j'$. The Hermiticity conditions for
the Hamiltonian (\ref{11}) and its negative-parity analog read
\begin{equation}
x_{j_{1}j_{2}}^{(+)}=(-)^{j_{1}-j_{2}}(x_{j_{2}j_{1}}^{(+)})^{\ast},
\quad {x'_{j_{1}'j_{2}'}}^{(+)}=
(-)^{j_{1}'-j_{2}'}({x'_{j_{2}'j_{1}'}}^{(+)})^{\ast}.  \label{14}
\end{equation}
The interaction with the octupole phonon that changes parity of the
particle is, analogously,
\[H^{(-)}=-\sum_{j_{1}j_{2}'m_{1}m_{2}\mu}y_{j_{1}j'_{2}}
\rho_{j_{1}m_{1};j_{2}'m_{2}}\]
\begin{equation}
\times\Bigl(f_{\mu}^{\dagger}+(-)^{\mu}f_{-\mu}\Bigr)
(-)^{\mu+j_{2}'-m_{2}}\left(\begin{array}{ccc}
j_{1} & 3 & j_{2}'\\
m_{1} & -\mu & -m_{2}\end{array}\right)+{\rm h.c.},     \label{15}
\end{equation}
Here we take the coordinate part of the phonon field, for example,
$(d_{\mu}^{\dagger}+(-)^{\mu}d_{-\mu}\Bigr)$, since it is
proportional to $1/\sqrt{\omega}$ in contrast to the momentum part
(with sign minus between the phonon creation and annihilation
operators) that is small, $\propto\sqrt{\omega}$, for a soft mode.

With the directions specified by the unpaired particle, the soft modes
acquire the condensate components in the ground state of the odd nucleus.
In the same way as shown by eq. (\ref{7}) we find
\begin{equation}
\langle d_{\mu}\rangle=\beta_{2}Y_{2\mu}^{\ast}({\bf n})=
\frac{1}{5\omega_{2}}\left[\sum_{j_{1}j_{2}}x_{j_{1}j_{2}}^{(+)}
\rho_{2}(j_{1}j_{2})+\sum_{j_{1}'j_{2}'}{x'_{j_{1}'j_{2}'}}^{(+)}
\rho_{2}(j_{1}'j_{2}')\right]Y_{2\mu}^{\ast}({\bf n}),    \label{16}
\end{equation}
\begin{equation}
\langle f_{\mu}\rangle=\beta_{3}Y_{3\mu}^{\ast}({\bf n})=
\frac{1}{7\omega_{3}}\left[\sum_{j_{1}j_{2}'}y_{j_{1}j_{2}'}
\rho_{3}(j_{1}j_{2}')+\sum_{j_{1}'j_{2}}y_{j_{1}'j_{2}}
\rho_{3}(j_{1}'j_{2})\right]Y_{3\mu}^{\ast}({\bf n}).   \label{17}
\end{equation}

\subsection{Density matrix of an unpaired particle}

The presence of phonon condensates self-consistently creates effective
deformation of the mean field for an unpaired particle. This is revealed
in the appearance of deformed components $\rho_{L\neq0}$ of the density
matrix (\ref{11}). The equations of motion for the single-particle
operators can be written as
\[[a_{jm},H]=\epsilon_{j}a_{jm}-\sum_{j_{1}m_{1}\mu}x_{j_{1}j}^{(+)}
\Bigl(d_{\mu}^{\dagger}+(-)^{\mu}d_{-\mu}\Bigr)(-)^{\mu+j-m}
\left(\begin{array}{ccc}
j_{1} & 2 & j\\
m_{1} & -\mu & m\end{array}\right)a_{j_{1}m_{1}}\]
\begin{equation}
-\sum_{j_{1}'m_{1}\mu}y_{j_{1}'j}\Bigl(f_{\mu}^{\dagger}+(-)^{\mu}f_{-\mu}\Bigr)
(-)^{\mu+j-m}\left(\begin{array}{ccc}
j_{1}' & 3 & j\\
m_{1} & -\mu & m\end{array}\right)a_{j_{1}'m_{1}}.     \label{18}
\end{equation}
A similar equation is valid for the operators $a_{j'm}$ related to
the negative parity levels with obvious substitutions
$x^{(+)}\rightarrow x'^{(+)}$ and $y_{j_{1}'j}\rightarrow
y_{jj_{1}'}$.

As now the odd particle is moving in the mean field of effective deformation,
its angular momentum $j$ is not conserved anymore but, owing to the
assumed axial symmetry, the angular momentum projection $\kappa$
onto the symmetry axis still characterizes the single-particle levels.
The appropriate transformation for the single-particle operators is
therefore
\begin{equation}
a_{jm}=\sum_{\kappa}D_{m\kappa}^{j}c_{j\kappa},        \label{19}
\end{equation}
where we use the matrix elements $D_{m\kappa}^{j}$ of finite
rotations. Such angular functions, after straightforward algebra,
appear in all terms of equations of motion (\ref{18}) justifying the
ansatz (\ref{19}). The new operators $c_{j\kappa}$ satisfy, instead
of (\ref{19}), the set of equations
\[[c_{j\kappa},H]=\epsilon_{j}c_{j\kappa}-2\beta_{2}
\sqrt{\frac{5}{4\pi}}\sum_{j_{1}}x_{j_{1}j}^{(+)}(-)^{j_{1}-\kappa}
\left(\begin{array}{ccc}
j_{1} & 2 & j\\
-\kappa & 0 & \kappa\end{array}\right)c_{j_{1}\kappa}\]
\begin{equation}
-2\beta_{3}\sqrt{\frac{7}{4\pi}}\sum_{j_{1}'}
y_{j_{1}'j}(-)^{j_{1}'-\kappa}\left(\begin{array}{ccc}
j_{1} & 3 & j\\
-\kappa & 0 & \kappa\end{array}\right)c_{j_{1}'\kappa},  \label{20}
\end{equation}
and analogously for the operators $c_{j'\kappa}$ for the levels
originally of negative parity. In the intrinsic frame parity is
mixed by the simultaneous presence of quadrupole and octupole
deformation {[}we used in eq. (\ref{20}) the static values
(\ref{16}) and (\ref{17})]. These coupled equations provide deformed
single-particle states and their energy spectrum. For the states
with one unpaired particle, the transformation (\ref{19}) is
normalized as
\begin{equation}
\sum_{jm}\langle a_{jm}^{\dagger}a_{jm}\rangle=\sum_{j\kappa}\langle
c_{j\kappa}^{\dagger}c_{j\kappa}\rangle=1.         \label{21}
\end{equation}

When the linear set of equations (\ref{20}) is solved and the
Nilsson-type single-particle orbitals with no definite parity are
found, we can return to the density matrix (\ref{11}) and make the
consideration self-consistent. From eq. (\ref{11}) we obtain
\begin{equation}
\rho_{L}(j_{1}j_{2})=\sqrt{4\pi(2L+1)}(-)^{L}\sum_{\kappa}
(-)^{j_{2}-\kappa}\left(\begin{array}{ccc}
j_{2} & j_{1} & L\\
-\kappa & \kappa & 0\end{array}\right) \langle
c_{j_{2}\kappa}^{\dagger}c_{j_{1}\kappa}\rangle.  \label{22}
\end{equation}
A similar expression is valid for $\rho_{L}(j_{1}'j_{2}')$ and for
the mixed-parity components of the density matrix
$\rho_{L}(j_{1}j_{2}')$.

\subsection{Schiff moment}

The operator $S_{1\nu}$ of the Schiff moment in the even nucleus
has a reduced matrix element $S^{\circ}\equiv(2||S_{1}||3)$ between
the low-lying $2^{+}$ and $3^{-}$ states. As mentioned earlier,
the dipole transitions between the states of the quadrupole and octupole
bands are empirically known to be enhanced in nuclei of our interest,
such as light radium and radon isotopes \citep{cocks97}. The collective
contribution to this operator can be written in terms of our phonon
variables as
\begin{equation}
S_{1\nu}=S^{\circ}\sum_{\mu\mu'}(-)^{\nu+\mu}
\left(\begin{array}{ccc}
1 & 2 & 3\\
-\nu & -\mu & \mu'\end{array}\right)
\Bigl(d_{\mu}^{\dagger}f_{\mu'}+
(-)^{\mu+\mu'}f_{-\mu'}^{\dagger}d_{-\mu}\Bigr).       \label{23}
\end{equation}
With the ground state expectation values of the effective
deformation parameters in the odd nucleus, eqs. (\ref{16}) and
(\ref{17}), this gives a rotational operator
\begin{equation}
\frac{S_{1\nu}}{S^{\circ}}=-\frac{1}{\sqrt{\pi}}\,\beta_{2}\beta_{3}
Y_{1\nu}^{\ast}                               \label{24}
\end{equation}
enhanced by small collective frequencies. As a result, we reduce the
whole problem to that of the {}``particle + rotor\char`\"{} type
\citep{leander84,SAF97}, where the static deformation is substituted
by the effective deformation coming from the soft quadrupole and
octupole modes of the spherical even core.

The observable Schiff moment in the laboratory frame can come only
from the explicitly acting ${\cal P}$- and ${\cal T}$-violating weak
interaction $W$ that creates an admixture of the states $|n\rangle$
of opposite parity to the ground state $|0\rangle$,
\begin{equation}
\langle{\bf S}\rangle=\sum_{n}\frac{W_{0n}{\bf S}_{n0}+{\bf
S}_{0n}W_{n0}}{E_{0}-E_{n}}.                      \label{25}
\end{equation}
The states $|n\rangle$ must have the same spin $I\neq0$ as the
ground state of the odd nucleus since $W$ is a rotational scalar.

\subsection{Simple model}

The structure of equations becomes clear in the simple model case of
two levels $j$ and $j'$ of opposite parity but the same magnitude
$j=j'$. The single-particle spherical energy levels are $\epsilon$
and $\epsilon'$, while the corresponding single-particle operators
in the intrinsic frame are $c_{\kappa}$ and $c_{\kappa}'$, where
$\kappa$ is the projection of the single-particle angular momentum
onto the symmetry axis. The analysis essentially follows the
consideration \citep{SAF97} developed for a statically deformed
nucleus with no symmetry with respect to the reflection in the
equatorial plane.

The effectively deformed core is presented by rotational states
$(D_{M'0}^{L})^{\ast}=\sqrt{4\pi/2L+1}\,Y_{LM'}$, where parity is
simply $(-)^{L}$. The particle orbitals are defined by the operators
$c_{\kappa}^{\dagger}$ for positive parity and
${c'_{\kappa}}^{\dagger}$ for negative parity multiplied by the
corresponding rotational functions $D_{m\kappa}^{j}$ which are
coupled with the core $D^{L}$-functions to the total angular
momentum $I$ and laboratory projection $M$ of the nucleus (the
body-fixed projection is $K=\kappa$). Due to the presence of
octupole deformation, the intrinsic particle states $|\kappa\rangle$
do not have certain parity. Their wave functions can be written as
\begin{equation}
|\kappa)=(\xi_{\kappa}c_{\kappa}^{\dagger}+
\eta_{\kappa}{c'_{\kappa}}^{\dagger})|0\rangle, \quad
\xi_{\kappa}^{2}+\eta_{\kappa}^{2}=1,             \label{26}
\end{equation}
and the amplitudes $\xi_{\kappa}$ and $\eta_{\kappa}$ satisfy the
secular equations with split energies $E_{\kappa}$,
\begin{equation}
[E_{\kappa}-\epsilon+b_{2}(\kappa)]\xi_{\kappa}+
b_{3}(\kappa)\eta_{\kappa}=0,                     \label{27}
\end{equation}
\begin{equation}
b_{3}(\kappa)\xi_{\kappa}+[E_{\kappa}-\epsilon'+
b_{2}(\kappa)]\eta_{\kappa}=0.                 \label{28}
\end{equation}
Here the effective deformation amplitudes  $b_{2}(\kappa)$ and
$b_{3}(\kappa)$ are given by
\begin{equation}
b_{2}(\kappa)=\sqrt{\frac{5}{4\pi}}\,\beta_{2}\,x_{2}, \quad
b_{3}(\kappa)=\sqrt{\frac{7}{4\pi}}\,\beta_{3}\,x_{3}.  \label{29}
\end{equation}
The energy spectrum of deformed orbitals is obviously
\begin{equation}
E_{\kappa}^{(\pm)}=\frac{\epsilon+\epsilon'}{2}\,-b_{2}(\kappa)\pm
\frac{1}{2}\,\sqrt{(\epsilon-\epsilon')^{2}+4b_{3}^{2}(\kappa)}.
                                           \label{30}
\end{equation}
The deformation parameters are found from eqs. (\ref{16}) and
(\ref{17}),
\begin{equation}
\beta_{L}=\frac{x_{L}}{(2L+1)\omega_{L}}\,\rho_{L},  \label{31}
\end{equation}
where $x_{2}=x_{jj}^{(+)}+{x'_{j'j'}}^{(+)}$ and $x_{3}=2y_{jj'}$.
The density matrix (\ref{22}) is now determined by the quantum
numbers $\kappa$ of the actually occupied pair of orbitals,
\begin{equation}
\rho_{L}(\kappa)=(-)^{L}\,\sqrt{4\pi(2L+1)}\,(-)^{j-\kappa}
\left(\begin{array}{ccc}
j & j & L\\
-\kappa & \kappa & 0\end{array}\right)z_{L}(\kappa),  \label{32}
\end{equation}
where
\begin{equation}
z_{2}(\kappa)=\xi_{\kappa}^{2}+\eta_{\kappa}^{2}=1,\quad
z_{3}(\kappa)=2\xi_{\kappa}\eta_{\kappa},        \label{33}
\end{equation}
and in the degenerate limit ($\epsilon=\epsilon'$) we have
$|\xi_{\kappa}|=|\eta_{\kappa}|=1/\sqrt{2},\;|z_{3}(\kappa)|=1$. Due
to time-reversal invariance, the same energy splitting (\ref{30})
occurs for the orbitals with the opposite sign of projection
$\kappa$.

\subsection{Parity doublets}

The single-particle intrinsic states $|\kappa)$ and $|-\kappa)$
found above, eq. (\ref{26}), should be combined into intrinsic orbitals
of certain parity,
\begin{equation}
|\psi_{\kappa;\pm})=\frac{1}{\sqrt{2}}\,
\Bigl[|\kappa)\pm(-)^{j-\kappa}|-\kappa)\Bigr].     \label{34}
\end{equation}
Since in the model each state $|\kappa)$ or $|-\kappa)$ is a
combination of orbitals based on mixed levels $j$ and $j'$, in fact
we have a quartet of relevant single-particle states or two parity
doublets with the same $|\kappa|$. If the matrix element of the weak
interaction between the single-particle spherical states $|jm)$ and
$|j'm)$ equals $w$, its analog in the intrinsic frame connecting the
partners of the doublets (\ref{34}) is $w\xi_{\kappa}\eta_{\kappa}$.
As it is clear from the symmetry arguments and seen from the set of
equations (\ref{27},\ref{28}), this product is proportional to the
octupole effective deformation $\beta_{3}$ and the component
$\rho_{3}$ of the single-particle density matrix. However, this
quantity becomes a constant in the exceptional case of degenerate
orbitals, eq. (\ref{32}). Then the maximum value of the
single-particle contribution of the weak interaction is, in this
model, just $|w|/2$.

The Schiff moment operator (\ref{24}) is responsible for the matrix
element between two rotational states with the same spin $I$ but
opposite parities. This geometric matrix element gives, as follows
also from general arguments \citep{SAF97,FKS84,FZ03},
\begin{equation}
\langle IM\kappa,\Pi|S_{10}|IM\kappa,-\Pi\rangle=
-\frac{1}{\sqrt{\pi}}\, S^{\circ}\beta_{2}\beta_{3}\,
\frac{M\kappa}{I(I+1)}.                       \label{35}
\end{equation}
 The final answer (\ref{25}) can be obtained multiplying the result
(\ref{35}) by $w\xi_{\kappa}\eta_{\kappa}/\Delta E_{I\kappa}$, where
the energy denominator is given by the splitting of rotational
partners of opposite parity. The collective enhancement for small
frequencies of the even core is explicitly present in eq.
(\ref{35}). The energy splitting $\Delta E_{I\kappa}$ is determined
mainly by the distance between the headbands of opposite parity that
is of the order of the difference of phonon frequencies
$\Delta\omega=|\omega_{2}-\omega_{3}|$. The effective moments of
inertia of bands in the odd nucleus built on the parity doublets are
expected to be close in heavy nuclei. For soft vibrations,
$\Delta\omega$ gives another enhancement factor.

\section{Numerical diagonalization of the model}

Here we show the results of the exact numerical solution of the two-level
model based not on the assumption of preexisting soft phonons but
on the full account of inter-particle interaction, including the primary
interaction responsible for the existence of soft modes. We will be
looking for the region of the parameter space that is favorable for
the enhancement effects.

\subsection{Nuclear Hamiltonian}

We consider, as above in Sec. 2.5, the single-particle space of two
spherical levels with the same $j=j'$ and opposite parity; their
spherical energies are taken as $\epsilon_{j}=0$ and
$\epsilon_{j'}=\epsilon$. The dynamics in the particle-particle
channel can be expressed in terms of the pair operators from the
same $j$-level,
\begin{equation}
P_{LM}=\frac{1}{\sqrt{2}}\,\sum_{mm'}C_{jm\,jm'}^{LM}\,a_{jm'}a_{jm},
\quad P_{LM}'=\frac{1}{\sqrt{2}}\,\sum_{mm'}C_{jm\, jm'}^{LM}\,
a_{j'm'}a_{j'm},                                  \label{36}
\end{equation}
where only even total momenta, $L=0,...,2j-1$, are allowed, and the
pair operator changing parity of the state,
\begin{equation}
R_{LM}=\sum_{mm'}C_{jm'\, jm}^{LM}\,a_{jm}a_{j'm'},   \label{37}
\end{equation}
where all values of the pair spin, $L=0,1,...,2j$, are possible. The
most general two-body Hamiltonian,
\begin{equation}
H_{{\rm int}}=H_{jj}+H_{j'j'}+H_{jj'}+V,         \label{38}
\end{equation}
includes the interactions inside each $j$-level,
\begin{equation}
H_{jj}=\sum_{LM}h_{L}P_{LM}^{\dagger}P_{LM}, \quad
H_{j'j'}=\sum_{LM}h_{L}'{P'_{LM}}^{\dagger}P_{LM}',  \label{39}
\end{equation}
the pair transfer between the levels,
\begin{equation}
H_{jj'}=\sum_{LM}g_{L}\Bigl(P_{LM}^{\dagger}P_{LM}'+
{P'_{LM}}^{\dagger}P_{LM}\Bigr),               \label{40}
\end{equation}
and, finally, scattering of the mixed pairs,
\begin{equation}
V=\frac{1}{4}\sum_{LM}V_{L}R_{LM}^{\dagger}R_{LM}.  \label{41}
\end{equation}

We specify the model choosing only the interactions associated with
the collective dynamics: pairing part $G>0$,
\begin{equation}
h_{0}=h'_{0}=g_{0}=-G;                        \label{42}
\end{equation}
quadrupole-quadrupole interaction in the particle-hole channel that,
after the angular momentum recoupling, corresponds to
\begin{equation}
h_{L}=h'_{L}=5\varkappa_{2}\, \left\{\begin{array}{ccc}
j & j & 2\\
j & j & L\end{array}\right\} ;                  \label{43}
\end{equation}
and octupole-octupole interaction that similarly gives rise to
\begin{equation}
V_{L}=7\varkappa_{3}\,\left\{ \begin{array}{ccc}
j & j & 3\\
j & j & L\end{array}\right\} .                 \label{44}
\end{equation}
Thus, the model Hamiltonian is characterized by four parameters,
$\epsilon,\, G,\,\varkappa_{2}$ and $\varkappa_{3}$. Attractive
forces correspond to positive $G$, $\varkappa_{2}$ and
$\varkappa_{3}$.

Also useful are the operators in the particle-hole channel,
\begin{equation}
Q_{\lambda\mu}^{\dagger}(j_{1}j_{2})=\sum_{m_{1}m_{2}}(-)^{j_{1}-m_{1}}
\left(\begin{array}{ccc}
j_{1} & \lambda & j_{2}\\
-m_{1} & \mu & m_{2}\end{array}\right)
a_{j_{1}m_{1}}^{\dagger}a_{j_{2}m_{2}},           \label{45}
\end{equation}
for different angular momenta $\lambda$ and all combinations of
$j_{1}$ and $j_{2}$. These operators have a symmetry property
\begin{equation}
Q_{\lambda\mu}^{\dagger}(j_{1}j_{2})=(-)^{j_{1}-j_{2}+\mu}
Q_{\lambda-\mu}(j_{2}j_{1}).                       \label{46}
\end{equation}
In these terms, our multipole-multipole interactions are
\begin{equation}
H^{(2)}=-\frac{5\varkappa_{2}}{2}\sum_{\mu}
\Bigl[Q_{2\mu}^{\dagger}(jj)Q_{2\mu}(jj)+Q_{2\mu}^{\dagger}(j'j')
Q_{2\mu}(j'j')\Bigr]                            \label{47}
\end{equation}
and
\begin{equation}
H^{(3)}=-\frac{7\varkappa_{3}}{2}\sum_{\mu}
\Bigl[Q_{3\mu}^{\dagger}(j'j)Q_{3\mu}(j'j)+Q_{3\mu}^{\dagger}(jj')
Q_{3\mu}(jj')\Bigr].                      \label{48}
\end{equation}
The Schiff moment operator is given by
\begin{equation}
S_{\mu}=-\sqrt{2j+1}\, s\Bigl[Q_{1\mu}(j'j)+Q_{1\mu}(jj')\Bigr],
                                             \label{49}
\end{equation}
while the weak interaction,
\begin{equation}
W=\sqrt{2j+1}\,w\Bigl[Q_{00}(j'j)+Q_{00}(jj')\Bigr],  \label{50}
\end{equation}
is characterized by one scalar $w$ for the transition between the
states $|jm)$ and $|j'm)$. As we are interested in the trends of the
expectation values of the Schiff moment in a function of nuclear
forces, the values of the coefficients $s$ and $w$ are irrelevant.
Below we set $w=1$ and $(j'm=j|S_{\mu=0}|jm=j)=\sqrt{j/(j+1)}$.

\subsection{Numerical results}

\begin{figure}
\includegraphics[clip,width=3.5in]{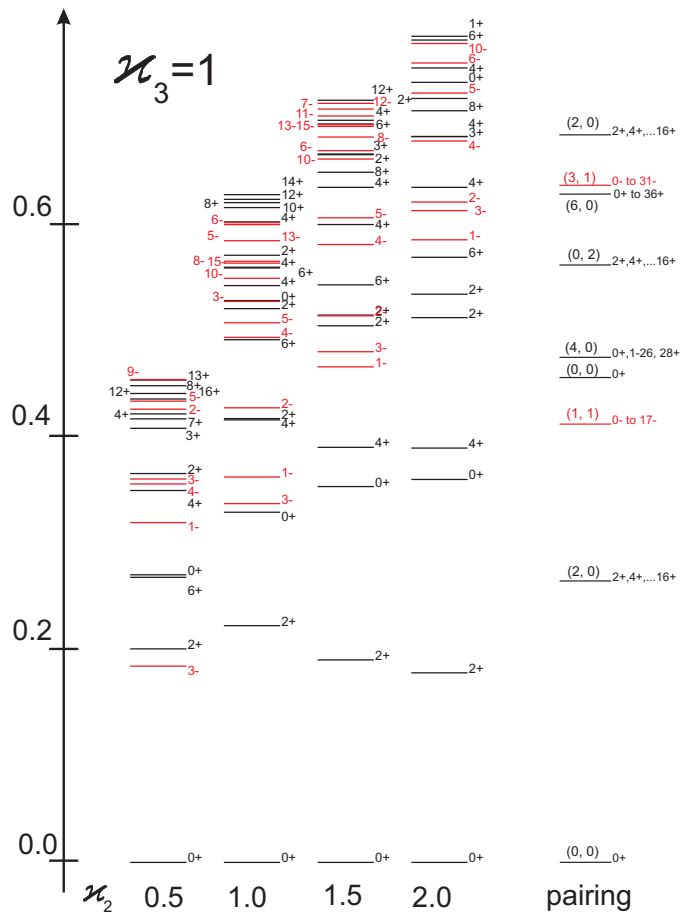}

\caption{The spectra of the 6-particle system for $\epsilon=0.2$ and
$G=0.2$. For the left four spectra, a fixed value of the octupole
strength, $\varkappa_{3}=1.0$, and a variable quadrupole interaction
parameter, $\varkappa_{2}$=0.5, 1, 1.5 and 2, are used. The most
right spectrum corresponds to the pairing Hamiltonian,
$\varkappa_{2}=\varkappa_{3}=0$, where seniorities $(s_{j},\,
s_{j'})$ are also marked. The negative parity states are shown in
red.\label{fig:spectrumN6o1}}

\end{figure}

Fig. \ref{fig:spectrumN6o1} shows the results for the energy
spectrum obtained in the exact diagonalization of the strong
Hamiltonian for 6 particles on two levels with $j=j'=17/2$. Here we
fix the octupole strength $\varkappa_{3}=1$ and the pairing strength
equal to the energy spacing between the single-particle levels,
$\epsilon=G=0.2$.

In the right column we show for comparison the spectrum for the pure
pairing interaction, $\varkappa_{2}=\varkappa_{3}=0$. As well known
from the analysis of the exact solution of the pairing problem
\citep{VBZ01}, partial seniority quantum numbers (or quasispins)
$s_{j}$ are conserved, and every eigenstate is characterized, apart
from the angular momentum quantum numbers $IM$, by the seniorities
$s_{j},s_{j'}$. The states with nonzero seniority are multiple
degenerate with respect to total spin. The states with the odd
seniority $s_{j'}$ have negative parity. The degenerate states with
$s_{j}=2,s_{j'}=0$ show a clear pairing energy gap that exceeds the
single-particle spacing $\epsilon$.

The first four columns show spectra in the presence of
multipole-multipole interactions for $\varkappa_{3}=1$ and different
values of $\varkappa_{2}$. The quadrupole collective state $2^{+}$
emerges within the pairing gap with energy only weakly changing as a
function of the quadrupole constant. At relatively weak quadrupole
strength, the collective low-lying $3^{-}$ state is also present.
However, as a consequence of our limited single-particle space, the
octupole and quadrupole branches essentially compete because, due to
their opposite parity, they require very different distortion
(polarization) of the quasiparticle vacuum.

\begin{figure}
\includegraphics[clip,width=3.5in]{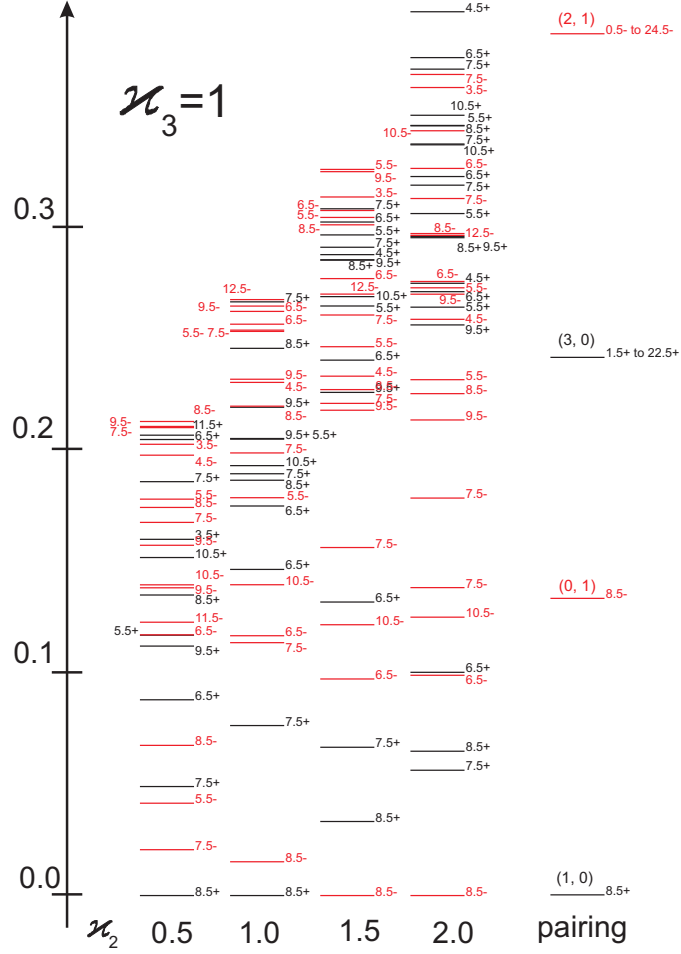}

\caption{Same as Fig. \ref{fig:spectrumN6o1} but for a 7-particle system.
\label{fig:spectrumN7o1}}

\end{figure}

The situation is different in the neighboring odd system ($N=7$),
Fig. \ref{fig:spectrumN7o1}, where the same Hamiltonian parameters
are used. In the case of the pairing only, we have two ``vacuum"
states with opposite parity and the same spin 17/2 which correspond
to a position of the odd particle on the $j$-level or on the
$j'$-level. This asymmetry sets into action the mechanisms
qualitatively described in Sec. 2. We see low-energy collective
excitations, which, in the region of approximately equal strength of
quadrupole and octupole branches, create parity doublets
$(17/2^{+},17/2^{-})$. Their minimum energy splitting, 0.15 at
$\varkappa_{2}=\varkappa_{3}=1$, is smaller than the single-particle
spacing.

\subsection{Degenerate case}

In the model with $j=j'$, the interactions are symmetric with
respect to the interchange ${\cal I}:(j\leftrightarrow j')$ of
levels of opposite parity. This operation commutes with spatial
inversion in an even system and anticommutes in the odd one, ${\cal
PI}=(-)^{N}{\cal IP}$, where $N$ is the number of particles. For
degenerate levels ($\epsilon=0$), ${\cal I}$ becomes the symmetry of
the strong Hamiltonian. But for an odd system, ${\cal I}$-symmetry
is incompatible with parity. Since the full ${\cal I}$-symmetry is
accidental, we still classify the nuclear eigenstates by parity
$\Pi$ but in this case two states of opposite parity are exactly
degenerate being related by the ${\cal I}$-operation. Thus, in the
odd system we have exact parity doublets, and all many-body states
are pairwise degenerate. In the even system, ${\cal P}$ and ${\cal
I}$ have common eigenstates with corresponding values $\pm1$.

In our definitions, both the weak interaction $W$ and the Schiff
moment operator ${\bf S}$ change parity but preserve ${\cal
I}$-symmetry. The transition matrix elements of ${\bf S}$ in the
even nucleus occur only within the class of states with the same
${\cal I}$. In the odd case we classify the stationary states (with
no weak interaction) by parity so that the new symmetry places no
constraints onto matrix elements of ${\bf S}$.

\begin{figure}
\includegraphics[clip,width=3.5in]{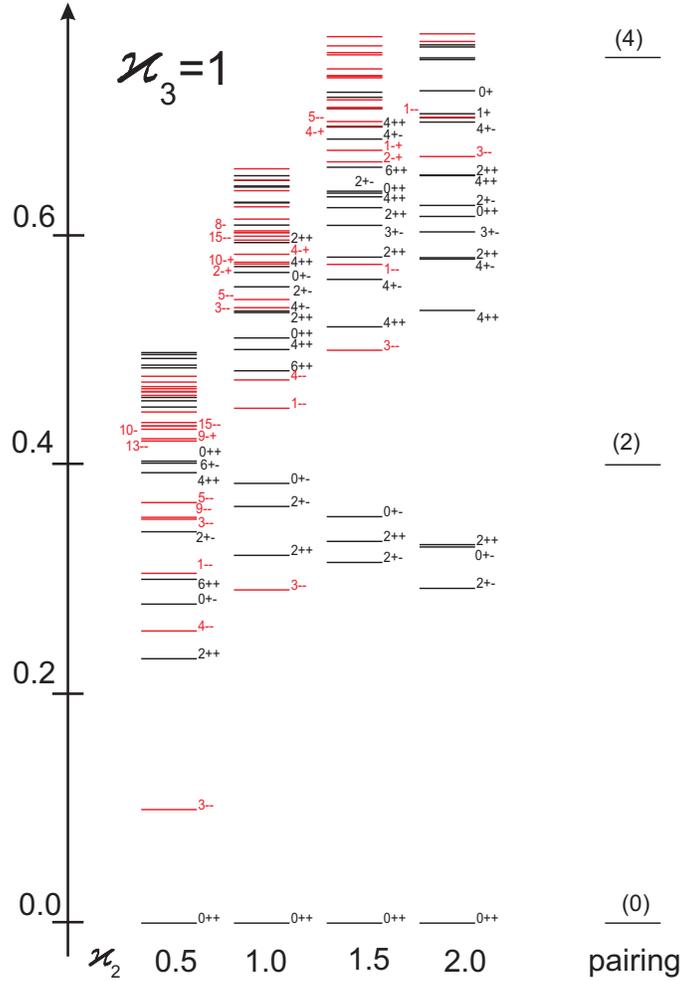}

\caption{The spectra of the 6-particle system with degenerate single
particle levels $\epsilon=0$; $G=0.2$ and $\varkappa_{3}=1.0$. The
set of values $\varkappa_{2}$=0.5, 1, 1.5 and 2 is used. The pure
pairing spectrum, $\varkappa_{2}=\varkappa_{3}=0$, with total
seniority marked is to the right. Most of the lowest states are
labeled as $J^{{\cal P}{\cal I}}$. The negative parity states are
shown in red.\label{fig:spectrumN6o1e0}}

\end{figure}

The spectrum of the exceptional case, $\epsilon=0$, is shown in Fig.
\ref{fig:spectrumN6o1e0} for $N=6$. In the pure pairing model (the
right column), the well known spin-degenerate seniority spectrum
($s=s_{j}+s_{j'}$) is given by
\begin{equation}
E(s)=-\frac{G}{2(2j+1)}\,(N-s)(4j+4-N-s),              \label{51}
\end{equation}
where even values of the seniority quantum number $s\neq0$ increase
energy through the Pauli blocking effect on quenched pairing. The
states for non-zero multipole-multipole forces are labeled by
quantum numbers $\Pi{\cal I}$. The situation with the lowest
collective states inside the gap is however not significantly
different from that observed for $\epsilon\neq0$. Again as
$\varkappa_{2}$ grows, we see the \textsl{destructive} influence of
the quadrupole mode onto the octupole mode; the latter is strongly
pushed up in energy. In our analytical consideration, Sec. 2.1, the
interplay of the quadrupole and octupole modes was
\textsl{constructive} (and this is what is seen in global
systematics and in xenon isotopes \citep{mueller06}). It follows
from eq. (\ref{4}) that the mutually supportive interaction of the
modes comes from the three-phonon ahnarmonicity, while the
conventional Hamiltonian of the type (\ref{38}) makes the
coexistence of the modes less probable, at least in the truncated
orbital space. We need to mention the crossing of the lowest
$2^{++}$ and $2^{+-}$ levels at $\varkappa_{2}=1.27$; then the
matrix element of the Schiff moment between the lowest states
$2^{+-}$ and $3^{--}$ does not vanish, in contrast to the matrix
element between $2^{++}$ and $3^{--}$.

\subsection{Schiff moment}

\begin{figure}
\includegraphics{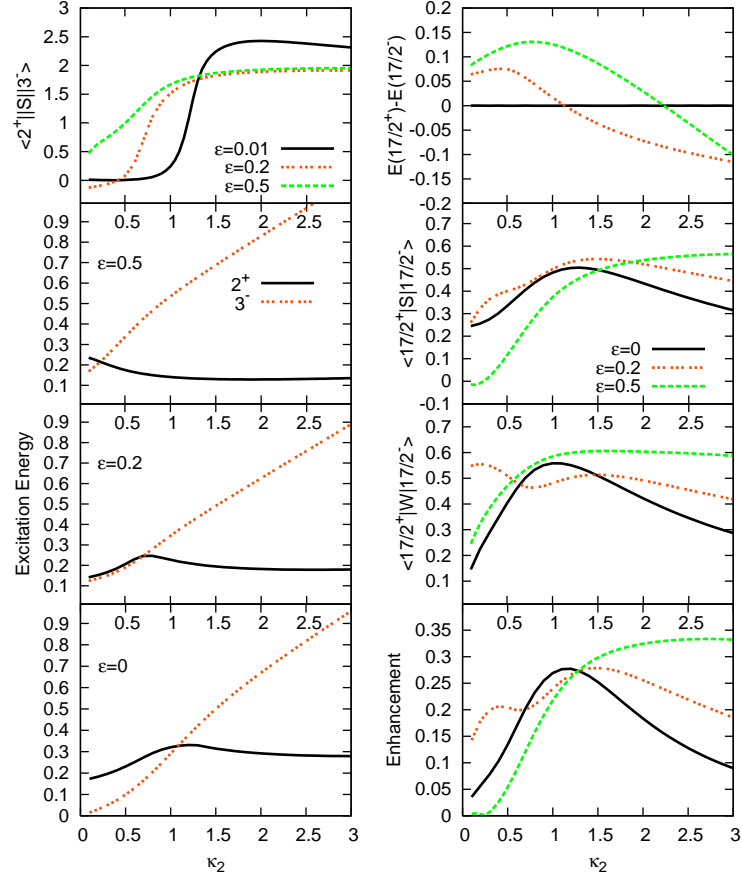}

\caption{The system of two $j=17/2$ levels is studied as a function
of quadrupole interaction strength $\varkappa_{2}$ at fixed values
$\varkappa_{3}=1$ and $G=0.2$. Three cases of single-particle
splitting are considered: $\epsilon=0$ (or very small value of 0.01
to remove abrupt numerical discontinuities), 0.2 and 0.5. The left
column is for the $N=6$ particle system. Three lower left panels
show the excitation energies of $2^{+}$ and $3^{-}$ levels for
different $\epsilon$ as labeled. The top left plot shows the reduced
matrix element of the Schiff operator between these states. The
7-particle system is depicted in the right column plots. The
splitting of the ground state doublet is on the right top followed
below by the plots of expectation values of the Schiff operator,
weak matrix element and their product. Three curves, solid (black),
dotted (red) and dashed (green), correspond to the same set of
single-particle spacings $\epsilon=0,\,0.2$ and 0.5, respectively.
\label{fig:S23}}

\end{figure}

Fig. \ref{fig:S23} shows the results for the matrix elements of the
Schiff moment (in the even and in the odd systems) and of the weak
interaction (in the odd system). In the even case (the left column,
$N=6$), the Schiff moment (the reduced matrix element between the
lowest {}``phonon\char`\"{} states $2^{+}$ and $3^{-}$ shown in the
upper left panel) strongly depends on the single-particle spacing
which was partly discussed above. For all values of $\epsilon$, the
maximum of this matrix element is reached in the region of
approximately equal strength $\varkappa_{2}$ and $\varkappa_{3}$.

The three remaining panels on the left show the excitation energies
of the lowest $2^{+}$ and $3^{-}$ states for different single-particle
spacings $\epsilon$. As $\epsilon$ and $\varkappa_{2}$ increase (at
fixed $\varkappa_{3}$), the destructive interaction of the modes pushes
the octupole phonon energy to high energy. However, at the reasonable
value $\epsilon=0.2$ that corresponds to the spectra shown earlier,
there is a region, where $\varkappa_{2}$ and $\varkappa_{3}$ are of the same
order and both resulting phonon frequencies are low.

The right column of Fig. \ref{fig:S23} gives the results for the odd
system, $N=7$, as a function of $\varkappa_{2}$. The upper panel
shows the splitting of the parity doublet $E(17/2^{+})-E(17/2^{-})$.
Both matrix elements, of the Schiff moment and of the weak
interaction, show a well pronounced maximum at
$\varkappa_{2}\sim\varkappa_{3}\sim1$ for not very large
single-particle spacing $\epsilon$. This is exactly what was the
purpose of the study. Of course, the precisely degenerate case,
$\epsilon=0$, is unphysical (in order to justify perturbation theory
(\ref{25}), the actual calculation here assumed a very small spacing
still exceeding the off-diagonal matrix element of the weak
interaction). However, a noticeable effect is present in the case of
$\epsilon=0.2$ as well.

\section{Conclusion}

We discussed the arguments in favor of the hypotheses of possible
enhancement of the nuclear Schiff moment, and therefore of the
effects of ${\cal P,T}$-violating weak interactions, in nuclei with
the simultaneous presence of soft collective modes of quadrupole and
octupole nature. If this idea is correct, the pool of possible
nuclei $-$ candidates for the successful search of the atomic EDM
$-$ would significantly broaden.

In our analytical model, we showed that in the odd neighbor of the
even spherical nucleus with soft vibrational modes the unpaired particle
leads to the spontaneously broken rotational symmetry. The arising
effective deformation can be described as a condensate of phonons.
The strength of the condensate is inversely proportional to the small
phonon frequencies in the even nucleus. This leads to the enhancement
of the intrinsic Schiff moment, in analogy to what was known for nuclei
with static quadrupole and octupole deformation. As a result, we obtain
the enhancement of the Schiff moment in the laboratory frame that
emerges due to the ${\cal P,T}$-violating weak interaction, whose
discovery is the primary goal of experimental and theoretical research
in this area.

The exact numerical study of a similar two-level model confirms the
existence of the region in the parameter space (single-particle
spectrum and the strengths of pairing and multipole-multipole
interactions) where the low-lying quadrupole and octupole modes
coexist, and where matrix elements of the weak interaction and of
the Schiff moment are indeed enhanced. The exact calculation does
not introduce any intrinsic frame and fully preserves rotational
symmetry and (in the absence of the weak interaction) parity of
stationary states. Although the model studied above had a limited
single-particle space and a small number of valence particles that
precludes the appearance of very strong collectivity, the presence
of the enhancement is visible.

We pointed out the difference between the analytical consideration
and the numerical model in the character of the interaction between
the quadrupole and octupole modes. The numerical study, being exact
in the framework of the two-level model, assumed the conventional
two-body Hamiltonian. Then the coexistence of the modes of
incompatible symmetry is suppressed by their destructive competition
for the available supply of single-particle excitations. The
analytical model assumed the strong anharmonic interaction between
the modes that leads to their mutual support and the appearance of
the quadrupole condensate in the state of the even nucleus with one
octupole phonon. Such cubic anharmonicities are rather weak if
considered in the next order of the random phase approximation built
on the two-body multipole-multipole interaction. However, currently
there is a broad discussion of the role of three-body forces, see
for example \citep{schwenk08}, in the nuclear structure beyond light
nuclei. The possible collective effects of three-body forces (in
this respect it does not really matter, bare or effectively induced
in nuclear matter) were briefly discussed, in relation to various
physical phenomena in nuclear physics, in \citep{Dallas06}. One of
the main new effects is the emergence of three-phonon anharmonicity.
The predicted positive correlation of quadrupole and octupole modes
is well seen in the chain of xenon isotopes \citep{mueller06}. Such
studies can be important in many questions of nuclear structure,
independently of the problem connected with weak interactions.

\section{Acknowledgments}

The work was supported by the NSF grant PHY-0555366 and by a grant
from the Binational Science Foundation USA-Israel.

\end{document}